\documentclass{jfm}

\usepackage{graphicx}
\usepackage{newtxtext}
\usepackage{newtxmath}
\usepackage{natbib}
\usepackage{hyperref}
\usepackage{soul}
\hypersetup{
    colorlinks = true,
    urlcolor   = blue,
    citecolor  = black,
}
\usepackage{amsmath,bm}
\usepackage{comment}

\newcommand{\bsm}[1]{\boldsymbol{#1} }

\newcommand{\RomanNumeralCaps}[1]
\linenumbers
\newcommand{\KSC}[1]{{\color{black}{#1}}}

\usepackage{xcolor}
\title{Formation of a strong negative wake behind a helical swimmer in a viscoelastic fluid}
\author{Shijian Wu\aff{1},
Tomas Solano\aff{2},
 Kourosh Shoele\aff{2}
 \and Hadi Mohammadigoushki\aff{1}\corresp{\email{hadi.moham@eng.famu.fsu.edu}}}

\affiliation{\aff{1}Department of Chemical and Biomedical Engineering, FAMU-FSU College of Engineering, Tallahassee, FL, 32310, USA
\aff{2}Department of Mechanical Engineering, FAMU-FSU College of Engineering, Tallahassee, FL, 32310, USA}

\begin{document}
\maketitle

\begin{abstract}
We investigate the effects of helical swimmer shape (i.e., helical pitch angle and tail thickness) on swimming dynamics in a constant viscosity viscoelastic (Boger) fluid via a combination of particle tracking velocimetry, particle image velocimetry and 3D simulations of the FENE-P model. The 3D printed helical swimmer is actuated in a magnetic field using a custom-built rotating Helmholtz coil. Our results indicate that increasing the swimmer tail thickness and pitch angle enhances the normalized swimming speed (i.e., ratio of swimming speed in the Boger fluid to that of the Newtonian fluid). Strikingly, unlike the Newtonian fluid, the viscoelastic flow around the swimmer is characterized by formation of a front-back flow asymmetry that is characterized by a strong negative wake \KSC{downstream of the swimmer's body}. Evidently, the strength of the negative wake is inversely proportional to the normalized swimming speed. Three-dimensional simulations of the swimmer with FENE-P model with conditions that match those of experiments, confirm formation of a similar front-back flow asymmetry around the swimmer. Finally, by developing an approximate force balance in the streamwise direction, we show that the contribution of polymer stresses in the interior region of the helix may provide a mechanism for swimming enhancement or diminution in the viscoelastic fluid.  
\end{abstract}
\section{Background}
Swimming of living organisms is a ubiquitous phenomenon in our world. Inspired by the pioneering work of G. I. Taylor, swimming in Newtonian fluids is well understood \citep{Lau09}. However, most organisms swim in complex fluids that contain polymers or proteins, which exhibit strong viscoelasticity. Prime examples include, swimming of sperms in cervical mucus for fertilization~\citep{Sua06}. Therefore, much attention has recently \KSC {been} devoted to understanding of the mechanisms of locomotion in non-Newtonian polymeric fluids. In particular, primary consideration has been given to two important non-Newtonian features including shear thinning~ \citep[e.g.][]{Zen17,Ard15} and/or elasticity~ \citep[e.g.][]{Lau07,Ter10,FuH07,Das13,Gom17,Liu11}.\par
Unlike shear-thinning, which is known to enhance the swimming speed in viscous fluids~\KSC{\citep{Zen17,Ard15}}, the impacts of elasticity on swimming dynamics in viscoelastic fluids \KSC{is a complex problem that depends on several parameters (see details below)}. Pioneering theoretical studies on an infinite sheet with small amplitude waves in 2D~\citep{Lau07} and 3D filaments~\citep{FuH07} have demonstrated that elasticity hinders the swimming speed. Lauga suggested the following analytical solution for the normalized swimming speed: ${U}/{U_{N}} = ({1+\beta De^{2}})/({1+De^{2}})$, where $U$, $U_{N}$, and $\beta$ denote the swimming speed in the viscoelastic fluid, swimming speed in the Newtonian fluid and the viscosity ratio defined as the ratio of solvent viscosity to zero-shear-rate viscosity of the viscoelastic fluid. Deborah number is defined as $De = \lambda \Omega$, where $\lambda$ and $\Omega$ are \KSC{longest} relaxation time of the polymer and the beating frequency of the swimmer, respectively. \KSC{For an undulatory swimmer with a large wave amplitude, 2D numerical simulations of~\citep{Ter10} showed that elasticity can enhance the swimming speed. A recent simulation study of E. coli motion in a polymer solution revealed that polymer molecules might migrate away from the bacteria surface leading to an apparent slip that may in turn cause an enhancement in the swimming speed~\citep{Zot19}.}\par
\KSC {In addition to theoretical studies, the effects of elasticity on swimming dynamics have been investigated in experiments.} Shen and Arratia studied swimming of a model worm (\emph{Caenorhabditis elegans}) in a viscoelastic fluid based on carboxymethylcellulose (CMC) in water, and showed that elasticity suppresses the swimming speed of C-elegans ($U/U_{N}<1$)~\citep{She11}, consistent with the asymptotic analysis of Lauga~\citep{Lau07}. Dasgupta et al.~\citep{Das13} noted swimming enhancement for a rotational Taylor-sheet in a Boger fluid based on polyacrylamide (PAA) in corn syrup/water. Moreover, Liu and co-workers showed that elasticity enhances the swimming speed ($U/U_{N}>1$) of a rigid helical (corkscrew) swimmer in a Boger fluid~\citep{Liu11}. Subsequently, other researchers performed experiments on three swimmers; i) a corkscrew rigid swimmer, ii) a 3D flexible swimmer and iii) a 2D flexible swimmer actuated by an external magnetic field in a Boger fluid based on PAA in corn syrup and water~\citep{God15,Esp13}. These authors showed that while elasticity does not affect the swimming speed of the rigid helical swimmer, it decreases the swimming speed of the 3D flexible swimmer, but, increases the swimming speed of the planar (2D) swimmer. \KSC{Patteson et al.~\citep{Patt15} reported swimming speed enhancement for an E. coli in viscoelastic polymeric solutions based on CMC. Additionally, these authors showed that E. coli follows a straighter trajectory in polymeric fluids compared to the Newtonian medium.} More recently, Qu $\&$ Breuer observed swimming speed enhancement for an E. coil in viscoelastic fluids and argued that swimming speed enhancement is predominantly caused by shear thinning, and normal stresses do not significantly impact the swimming speed~\citep{Qu20} . \par

More recent theoretical studies have suggested that locomotion dynamics in viscoelastic fluids is a complex problem that not only depends on rheological properties of the medium, but also the swimmer shape and its actuation mode~\citep{Spa13,Ril15,Tho14,Qu20}. In particular \cite{Spa13} investigated swimming dynamics of an infinitely long helical swimmer in an Oldroyd-B (OB) constitutive fluid model. \cite{Spa13} demonstrated that elasticity increases the swimming speed of helical swimmers with sufficiently large pitch angle (e.g., $\psi>\pi/5$). Additionally, they showed that at a fixed pitch angle, increasing the tail thickness of the helical swimmer reduces the normalized swimming speed. \KSC{ For a more detailed analysis of the literature on swimming dynamics in viscoelastic fluids, readers are referred to recently published review papers ~\citep{Li2021,PATTESON201686}}. \par

While the above theoretical analysis of Spagnolie and co-workers have provided valuable insights on helical swimming dynamics in viscoelastic polymer solutions, these predictions have not been systematically verified in experiments. Additionally, pertaining experimental studies on helical swimming have mostly characterized the swimming speed and therefore, not much information about the structure of the flow around the swimmer is available. The detailed form of flow structure around the swimmer provides crucial information that may shed more light on the origin of swimming enhancement or retardation in viscoelastic fluids. Finally, \KSC {the }majority of theoretical studies of swimming in viscoelastic fluids including those of \citep{Spa13} have used the OB fluid model. Although \KSC {the} OB model is very attractive for viscoelastic polymer solutions, \KSC{it predicts that polymer chains can be extended indefinitely, which is aphysical and therefore, may be expected to provide inadequate description of the flow of polymeric chains around a swimmer.} Flow around the swimmer is highly complex and non-viscometric with strong evidence for formation of extensional flows. \KSC {For example, Shen \& Arratia reported formation of a hyperbolic point near swimming C-elegans~\citep{She11}. Additionaly, 3D simulations of a C-elegan in a Gisekus fluid model demonstrated that the extensional flow is strong near the tips of the swimmer~\citep{Bin19}. In fact, it has been recently shown that for a model squirmer with swirling motion at relatively weak viscoelasticity (i.e., small Weissenberg numbers, $Wi <1$), the OB model produces a singularity in elastic stresses near the poles of the squirmer, where the extensional flow is dominant~\citep{Hou21}. This singularity was removed by use of more advanced constitutive models such as the finitely extensible nonlinear elastic model with Peterlin approximation (FENE-P) and Giesekus models~\citep{Hou21}. Our flow visualization experiments (see details below) in the viscoelastic fluid also revealed formation of strong extensional flows around a helical swimmer. Therefore, to provide a more accurate description of the flow of viscoelastic fluids in complex flows such as those found around a helical swimmer we will use the FENE-P model.}\par
  
The main objective of this paper is to provide a systematic understanding of the effects of helical swimmer shape on swimming dynamics in viscoelastic fluids. In particular, we will assess how swimming dynamics and form of flow structure around a 3D printed swimmer are affected by variation of the helical pitch angle and the tail thickness in a model Boger fluid. We will employ a combination of particle tracking velocimetry and particle image velocimetry to measure the velocity of the swimmer and the detailed form of the flow structure around the swimmer. Finally, we will compare our experimental results with three dimensional simulations of the FENE-P model. 
\section{Experimental and Computational Approaches}
	\subsection{Experiments}
\begin{figure} 
\centering
	\includegraphics[width=0.99\textwidth]{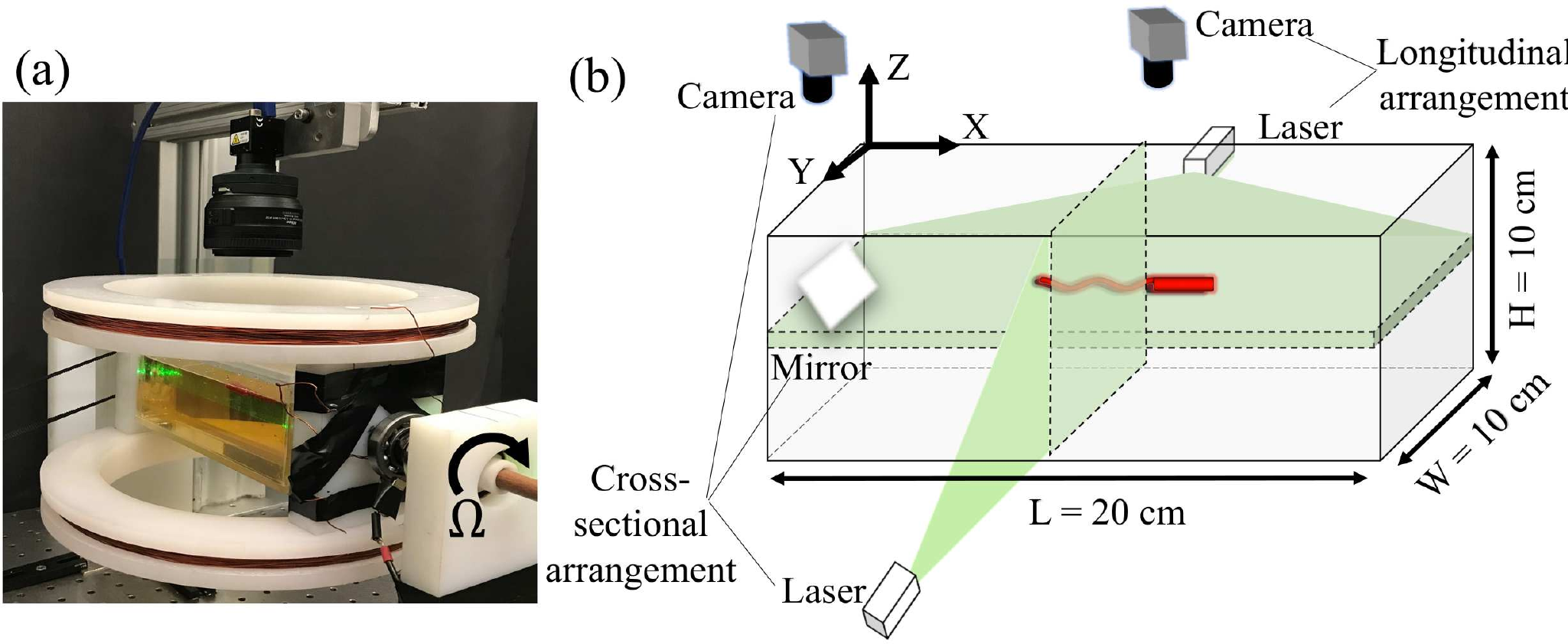}
	\caption{\small (a) The custom-made rotating Helmholtz-coil used for this study. (b) A schematic of the fluid chamber and different laser arrangements used to obtain the detailed form of the flow structure around the swimmer. The swimmer is moving with a constant velocity in the x-direction. The tail length $L_T$, head length $L_h$ and head diameter $D$ are held fixed as 3.7cm, 2.3cm and 4.5mm, respectively. The tail pitch angle is varied as: $\psi = 12.5^{\circ}-66.5 ^{\circ}$ and three tail thicknesses of $d =$ 1.1, 1.5 and 2.1mm are considered.}
	\label{fig1}
\end{figure}

Figure~\ref{fig1} shows the experimental device used for this study. Experiments are performed using a range of 3D printed swimmers with a helical tail and a cylindrical head. A small cylindrical magnet is embedded in swimmers’ head and actuated via a uniform rotating magnetic field imposed by the custom-made rotating Helmholtz coil, which is similar to the setup used before~\citep{God12}. \KSC{In these experiments the helical swimmer is placed in a chamber which is filled with the fluid. The chamber is positioned at the middle of the Helmholtz coil
and a uniform magnetic field is generated in the y-direction by passing a DC current through the wires that are wrapped around the two rotating circular features.} Time-resolved particle tracking velocimetry (PTV) is performed to measure the swimming speed. \KSC{More details about PTV analysis can be found in our previous publications~\citep{Wu18,WU2019}}. Additionally, fluids are seeded by a small amount (40ppm) of seeding particles \KSC{(model 110PB provided by Potters Industries LLC) to enable particle image velocimetry (PIV). This small amount of seeding particles does not change the rheology of the fluids used in this work. Flow visualization is performed in two different planes (X-Y and Y-Z; see Fig.~\ref{fig1}(b)) via a high-speed camera (model Phantom Miro-310) equipped with different macro lenses (Nikon). For the PIV analysis, we used an open source MATLAB code developed by Thielicke and Stamhuis~\citep{Thie14}.} \\

\KSC{Steady shear and the small amplitude oscillatory shear experiments are performed via a commercial MCR 302 Anton-Paar rheometer. Additionally, the transient extensional rheological properties of the viscoelastic fluid are characterized using a custom-made capillary breakup extensional rheometer (CaBER). More details about this device can be found in our previous publication~\citep{Om18}.}


\subsection{Governing Equations:} The conservation of mass and momentum are used to calculate for the transient flow response of the incompressible viscoelastic fluid:
\begin{subequations}
\label{equations}
\begin{align}
\label{eq:2.1}
\rho \left(\frac{\partial \bm{u}}{\partial t}{+}\bm{u}\boldsymbol{\cdot }\boldsymbol{\nabla }\bm{u}\right) = -\boldsymbol{\nabla }p +  \boldsymbol{\nabla } \boldsymbol{\cdot }\boldsymbol{\tau},~~~~~~~
\boldsymbol{\nabla \cdot} \bm{u} = 0,
\tag{2.1 a,b}
\end{align}
\end{subequations}	    

\noindent where $\mathbf{u}$, $p$ and $\boldsymbol{\tau}$ are \KSC {the} velocity vector, pressure and effective stress tensor, respectively. $\boldsymbol{\tau}=\bsm{\tau}_s+\bsm{\tau}_e$ is the summation of solvent viscous stress and polymer stress. The solvent stress tensor is $\boldsymbol{\tau}_s\,=\,\mu_s(\boldsymbol{\nabla}\mathbf{u}+\boldsymbol{\nabla}\mathbf{u}^T)$, where $\mu_s$ is the solvent viscosity. The polymer stress, $\boldsymbol{\tau}_e$, is calculated via FENE-P constitutive model~\citep{bird1980polymer,purnode1998polymer} as:
\begin{equation}
\bsm{\tau}_e\,+\,\frac{\lambda}{f} \overset{\nabla}{\bsm{\tau}_e}\,+\, \frac{D}{Dt}\left(\frac{1}{f}\right) \left(\lambda \bsm{\tau}_e+l\mu_p \textbf{I}\right)\,=\,\frac{l\, \mu_p}{f}\left[\boldsymbol{\nabla}\mathbf{u}+(\boldsymbol{\nabla}\mathbf{u})^T\right]\label{eqV1}
\end{equation}
where $ \lambda$ is the polymer relaxation time, $\mu_p$ \KSC {is the} polymer viscosity, $D/Dt$ is the material derivative, and $l=L^2/(L^2-3)$ where  	$L$ measures the extensibility of the polymer chains ($L^2=3R^2/R_0^2$ with $R$ being the maximum length of polymer chain and $R_0$ being the equilibrium length of the chain). The function $f$ is defined as $f=l+\frac{\lambda}{L^2\,\mu_p}\text{tr}(\bsm{\tau}_e)$. Finally, $\overset{\nabla}{\,}$ represents the \emph{upper-convected derivative} given by,
\begin{equation}
    \overset{\tiny{\nabla}}{\bsm{\tau}_e}= \frac{D\bsm{\tau}_e}{Dt}\,-\,\bsm{\tau}_e \bsm{\cdot} \bsm{\nabla}\mathbf{u}\,-\, \bsm{\nabla}\mathbf{u}^T \bsm{\cdot} \bsm{\tau}_e. \label{eqV2}
\end{equation}
The first-order positivity preserving method \citep{stewart2008improved} is used to solve Eqs~\ref{eqV1} and \ref{eqV2}). The simulations are performed using the hierarchy grids from dynamic adaptive mesh refinement technique (AMR)  for computational efficiency \citep{sussman1999adaptive}. \KSC{In particular, the computational setup consists of 4 levels of refinement covering a Cartesian domain of $10L_h \times 3L_h \times 3L_h$, where $L_h$ is the length of the swimmer head. Further increase of the domain size has a negligible effect on the results (less than 1\%). The most refined grid is uniform with the resolution of $D/22$  placed around the body at $-D$ to $D+L_h+L_t$ in the streamwise direction and $-1.5D$ to $1.5D$ in the lateral directions from the center of the upfront point of the swimmer's head. Here $L_t$ and $D$ refer to the tail length and the head cross sectional diameter. A mesh refinement study is performed to ensure a sufficient grid resolution wherein it is found that the changes in hydrodynamic forces  are less than 2\% with further refinement. The far-field boundary condition is used at all boundaries and a fixed time step of $\Delta t=0.001 \Omega^{-1}$ is employed for all simulations. }\\

\KSC{All simulations are performed at vanishingly small Reynolds numbers. The Reynolds number is defined as $Re_{\Omega}=\rho D^2 \Omega /\mu_s$ and its value is kept fixed at $Re_{\Omega}=0.000625$ in simulations, consistent with the experiments with a maximum Reynolds number (corresponding to the maximum rotation rate) of $Re_{\Omega}= 0.0125$.} The  details of the numerical techniques used for solving the momentum and mass conservation equations are given in~\citep{vahab2021fluid}. The model validation for related problems are presented in \citep{vahab2021fluid,stewart2008improved,ohta2019three}.

\section{Results}

\subsection{Fluid Characterization}

\begin{figure}
	\centering\includegraphics[width=1\textwidth]{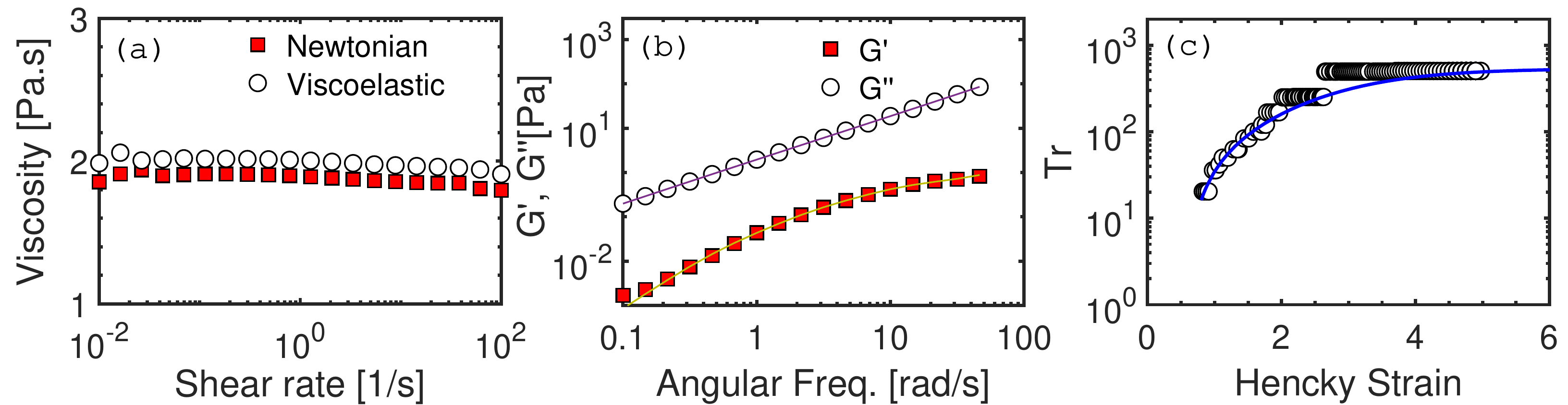}
	\caption{\small (a) Steady state shear viscosity as a function of shear rate for different fluids. (b) Elastic ($G'$; empty symbols) and loss moduli ($G''$; filled symbols) measured for the Boger fluid at room temperature. In part (b), the solid lines denote the best fit to a four-mode Maxwell model, which yields a longest relaxation time of $\lambda = 1.52~$sec. (c) Transient Trouton ratio as a function of \KSC{total} accumulated strain (Hencky strain) for experiments (symbols) along with the best fit to the FENE model (curve).} 
	\label{fig2}
\end{figure}

Two model fluids have been characterized and used in this study; a Newtonian fluid based on corn syrup/water (95 wt$\%$/5 wt$\%$) and a viscoelastic fluid that contains 150 ppm Polyacrylamide (PAA) in corn syrup/water (95wt$\%$/5wt$\%$). As shown in Fig.~\ref{fig2}(a), the viscoelastic fluid exhibits a constant viscosity over a broad range of applied shear rates\KSC{, and a viscosity ratio (i.e., the ratio of the solvent viscosity to that of the total solution viscosity) of $\beta$ = 0.9 is obtained.} Additionally, the results of small amplitude oscillatory shear experiments (Fig.~\ref{fig2}(b)) are fitted to a four-mode Maxwell model, which yields \KSC{the longest relaxation time of }$\lambda$ = 1.52 sec. Moreover, Fig.~\ref{fig2}(c) shows the \KSC{transient Trouton ratio as a function of Hencky (total accumulated) strain measured via CaBER for the viscoelastic fluid. The Trouton ratio is defined as $Tr = \eta_E/\eta_0$, where $\eta_E$ and $\eta_0$ denote the apparent extensional viscosity and the zero shear rate viscosity of the polymeric fluid. More details on how to obtain the transient extensional viscosity and the total accumulated strain can be found in our prior publications~\citep{Omid19,Om18}. Included in this figure is also the best fit to the FENE model, which yields a finite extensibility of $L^2 \approx 9$. } The finite extensibility of the polymer chains is used for simulations of the swimming in the FENE-P fluid model. \par 

\subsection{Swimming Speed}
\begin{figure}
		\centering
		\includegraphics[width=0.8\textwidth]{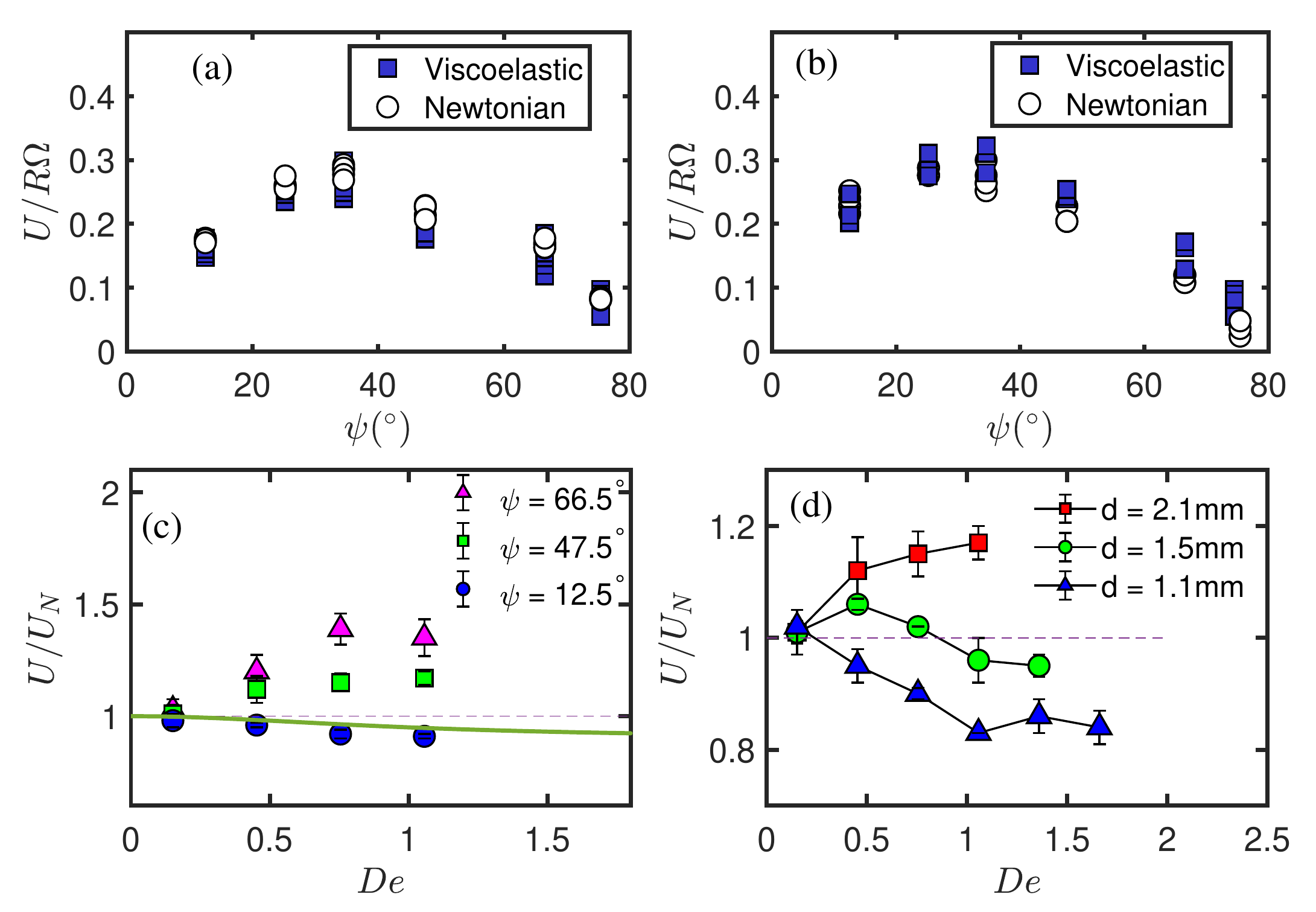}

		\caption{\small \KSC{Dimensionless swimming speed ($U/R\Omega$) as a function of pitch angle ($\psi$) measured in the Newtonian and the viscoelastic fluid for tail thickness of $d = 1.1$ mm (a) and $d = 2.1$ mm (b).} Normalized swimming speed ($U/U_N$) as a function of Deborah number for (a) various pitch angles at a fixed tail thickness ($d=2.1~ mm$) and (b) various tail thickness and a fixed pitch angle ($\psi=47.5^\circ$). In part (a) the experimental data for $\psi=12.5^\circ$ are compared to the analytical solution of Lauga (continuous line;~\citep{Lau07}).}
		\label{fig3}
\end{figure}

\KSC{Fig.~\ref{fig3} (a,b) show the dimensionless swimming speed ($U/R\Omega$) as a function of pitch angle in the Newtonian and the viscoelastic fluid for different tail thicknesses ($d = 1.1 mm$ in Fig.~\ref{fig3} (a) and $d = 2.1 mm$ for Fig.~\ref{fig3} (b)). For both fluids, we report a similar optimal pitch angle ($\psi \approx 35^{\circ}$) that generates the maximum swimming speed. In addition, for the smaller tail thickness ($d = 1.1$ mm) the swimming speed is smaller in the viscoelastic fluid than the Newtonian counterpart. At a larger tail thickness ($d = 2.1$ mm), the swimming speed in the viscoelastic fluid is similar to that of the Newtonian fluid for sufficiently small pitch angles ($\psi \leq 25.5 ^{\circ}$), and at larger pitch angles, the swimming speeds in the viscoelastic fluid are larger than those measured in the Newtonian fluid. Note, that multiple symbols at a given tail pitch angle in Fig.~\ref{fig3} (a,b) correspond to different rotation rates (or equivalently different $De$ numbers for experiments in the viscoelastic fluid). }\par

Fig.~\ref{fig3}(c) shows the normalized swimming velocity ($U/U_N$) as a function of Deborah number $De$ for swimmers with a fixed tail thickness ($d = 2.1 $mm) and different tail pitch angles. For the lowest pitch angle ($\psi = 12.5^\circ$), the normalized swimming speed is around unity for $De \approx 0.1$, and as the elasticity (or $De$ number) increases, the normalized swimming speed remains below unity. Included in this figure is the prediction of the asymptotic model of Lauga for low amplitude swimmers, which agrees well with the experimental results. \KSC{Note that the asymptotic model of Lauga has been developed for a low-amplitude two-dimensional swimmer with no head, while all of our 3D printed swimmers have a fixed cylindrical head. Nevertheless, the agreement between our experiments and Lauga's model may imply that at low amplitudes, such geometrical differences may not produce significant differences in swimming speeds.} As the pitch angle increases, the normalized swimming speed increases beyond unity. The latter result is consistent with the predictions of Spagnolie et al.~\citep{Spa13} for a rigid helix with an infinite length in an Oldroyd-B fluid model. However, note that, the extent of speed enhancement in these experiments is larger than what is typically predicted by Spagnolie et al. ($\approx 5\%$)~\citep{Spa13}. \KSC{The difference in extent of swimming enhancement could presumably be due to geometrical differences between our finite-size 3D printed swimmers and those used by Spagnolie and co-workers (a headless infinite helix). It is possible that at sufficiently large pitch angles, the presence of a cylindrical head and/or the finite size of the 3D printed swimmers have contributed to additional swimming speed enhancement in experiments. This hypothesis will be tested in our future work. Note that in some experiments (including the results of Fig.~\ref{fig3}(c)), we cannot access Deborah numbers beyond 1. If the rotation rate is high, the small magnet in the swimmer head can no longer synchronize with the magnetic field of the Helmholtz coil and step out. This step out frequency has been noted in previous studies by Zenit and co-workers (see e.g., Figure~(3) in \citep{God12}).} \par 

Fig.~\ref{fig3}(d) illustrates the normalized swimming speed as a function of $De$ for a fixed pitch angle and different tail thicknesses. Interestingly, at low tail thickness, the normalized swimming speed is below unity and as the tail thickness increases, the normalized swimming speed increases. These results are in apparent disagreement with the existing predictions of Spagnolie and co-workers, wherein, increasing the helix thickness was shown to hinder the swimming in an Oldroyd-B fluid model~\citep{Spa13}. \KSC {As noted in the above, there are two geometrical differences between swimmers used in this study and that of Spagnolie and co-workers' that may contribute to the above discrepancy; presence of the head and the finite length of the 3D printed swimmers. In experiments of Fig.~\ref{fig3}(d), the size of the head is fixed. Therefore, the observed swimming enhancement as a result of increasing the tail thickness may not by affected by the presence of the head. However, the finite size of the swimmers used in this study may lead to strong extensional flows around the swimmer. These flow features around the swimmer in experiments may have contributed into the different trends seen in experiments and simulations of Spagnolie and co-workers.} In the biological environment, multiflagellated bacteria (e.g., H. pylori or E. Coli) increase the thickness of their tail by forming tight bundles of synchronized flagella in viscous and viscoelastic fluids~\citep{Martinez16,Ban13}. \KSC{Similarly, Breuer, Powers and co-workers showed that rotation of model flexible helices generates a flow that may cause flagella bundling in viscous fluids~\citep{Kim03}.} In fact, it has been shown that bundling of more flagella in different strains of H. pylori, enhances bacterium swimming speed~\citep{Martinez16}. The latter outcome is consistent with our results that increasing the tail thickness leads to swimming enhancement.   \par 
\KSC{ Zenit and co-workers have also studied swimming dynamics of a series of swimmers similar to our 3D printed swimmers (with a helical tail and a cylindrical head)~\citep{Ang2021,God15}. A direct comparison between our swimming speed results and those of Zenit and co-workers is not possible because not only the details of the swimmer geometry is different (head diameter, tail thickness, total length of the swimmer, proportion of the tail length to that of the head length) from our swimmers, but also the Boger fluids have a slightly different viscosity ratios ($\approx$ 0.92 and 1~\citep{Ang2021} or 0.95~\citep{God15}). Additionally, the tail of the swimmer in their experiments is made up of steel wire that may have a different Young modulus/bending stiffness than the polylactic acid that we used for our 3D printed swimmers. Tail flexibility is known to significantly impact the swimming dynamics in viscoelastic fluid~\citep{Ril15}.}


\subsection{Flow Structure Around the Swimmer}
\begin{figure}
		\centering
		\includegraphics[width=0.325\textwidth]{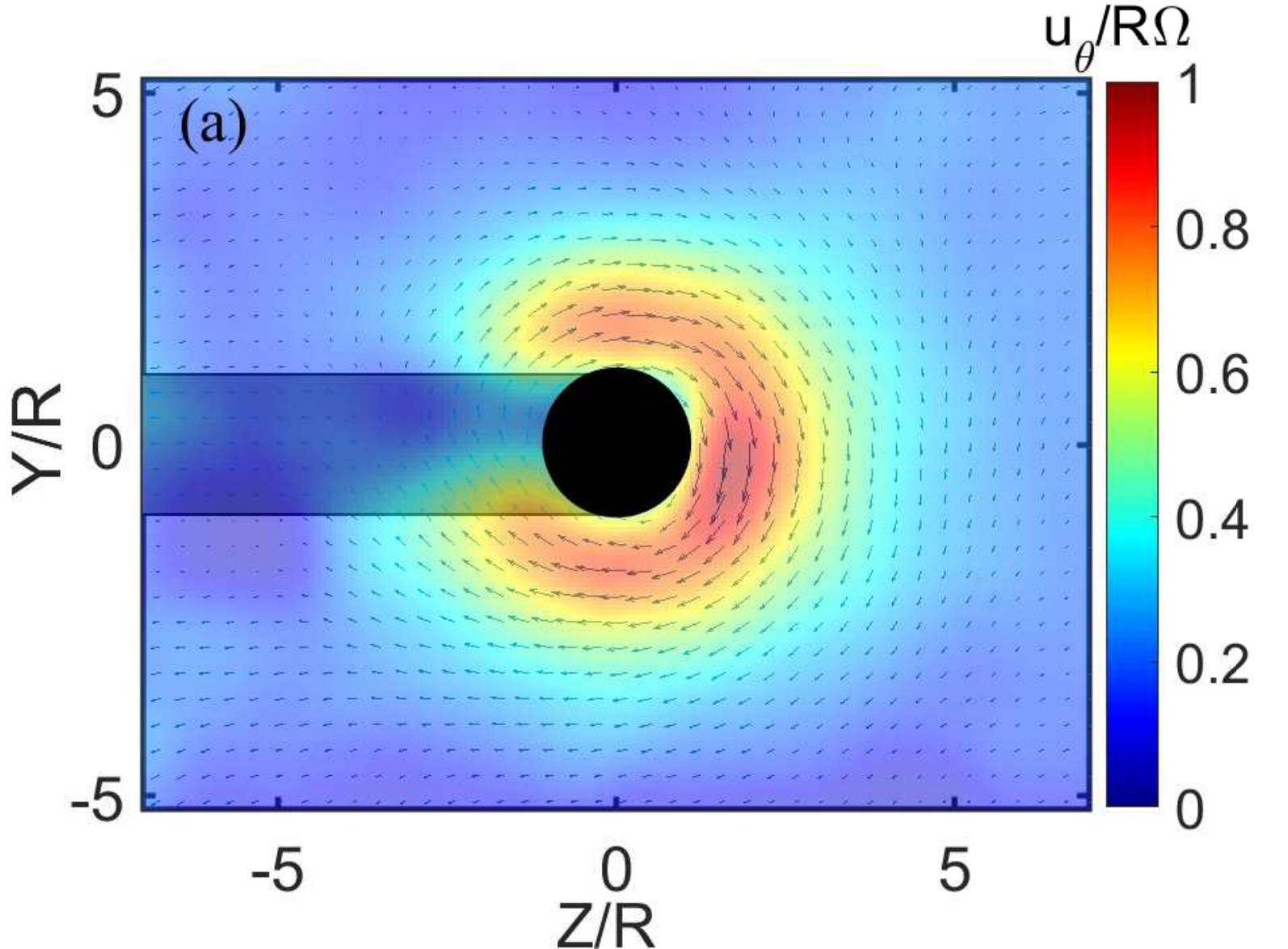}
		\includegraphics[width=0.325\textwidth]{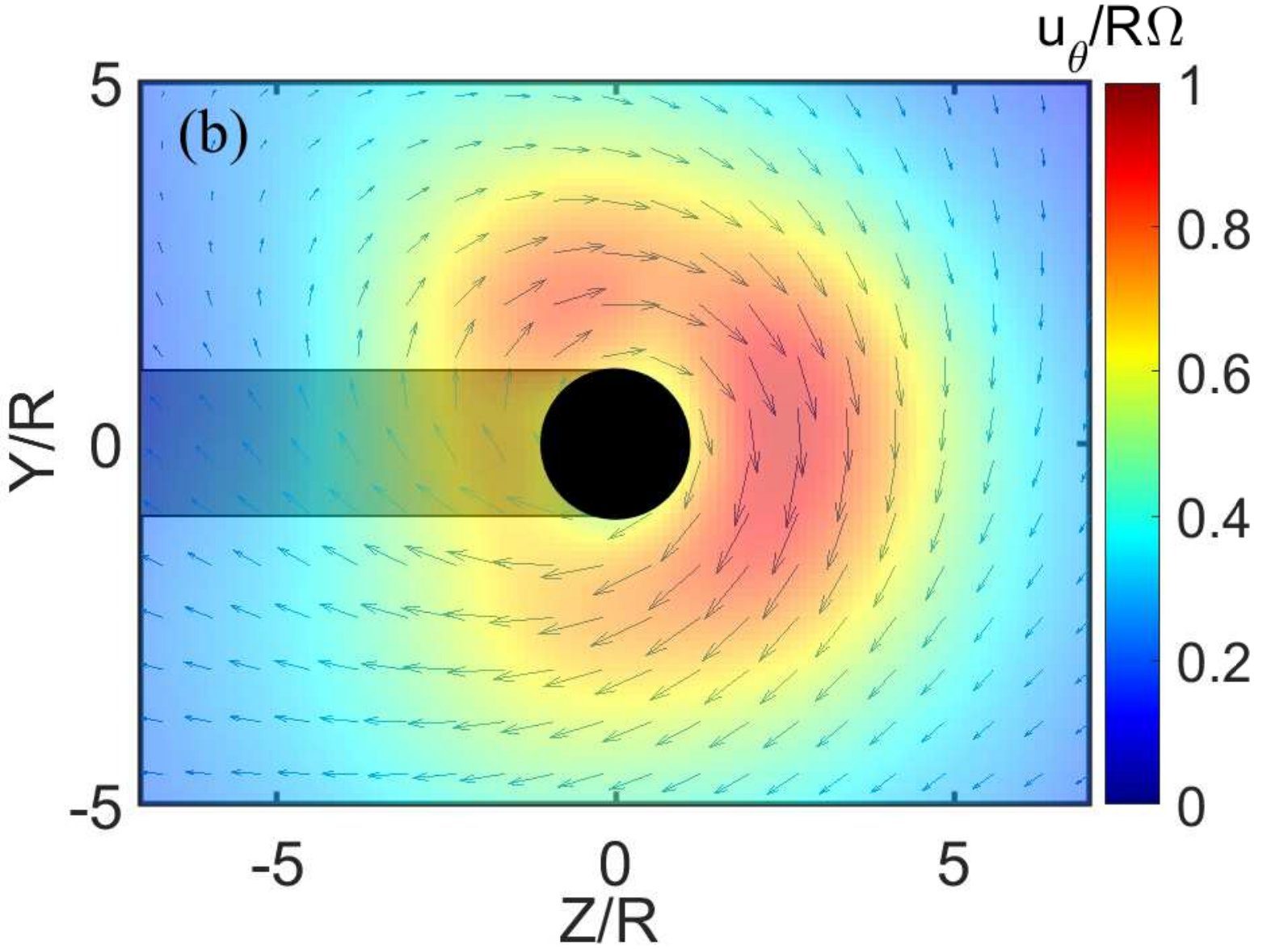}
		\includegraphics[width=0.33\textwidth]{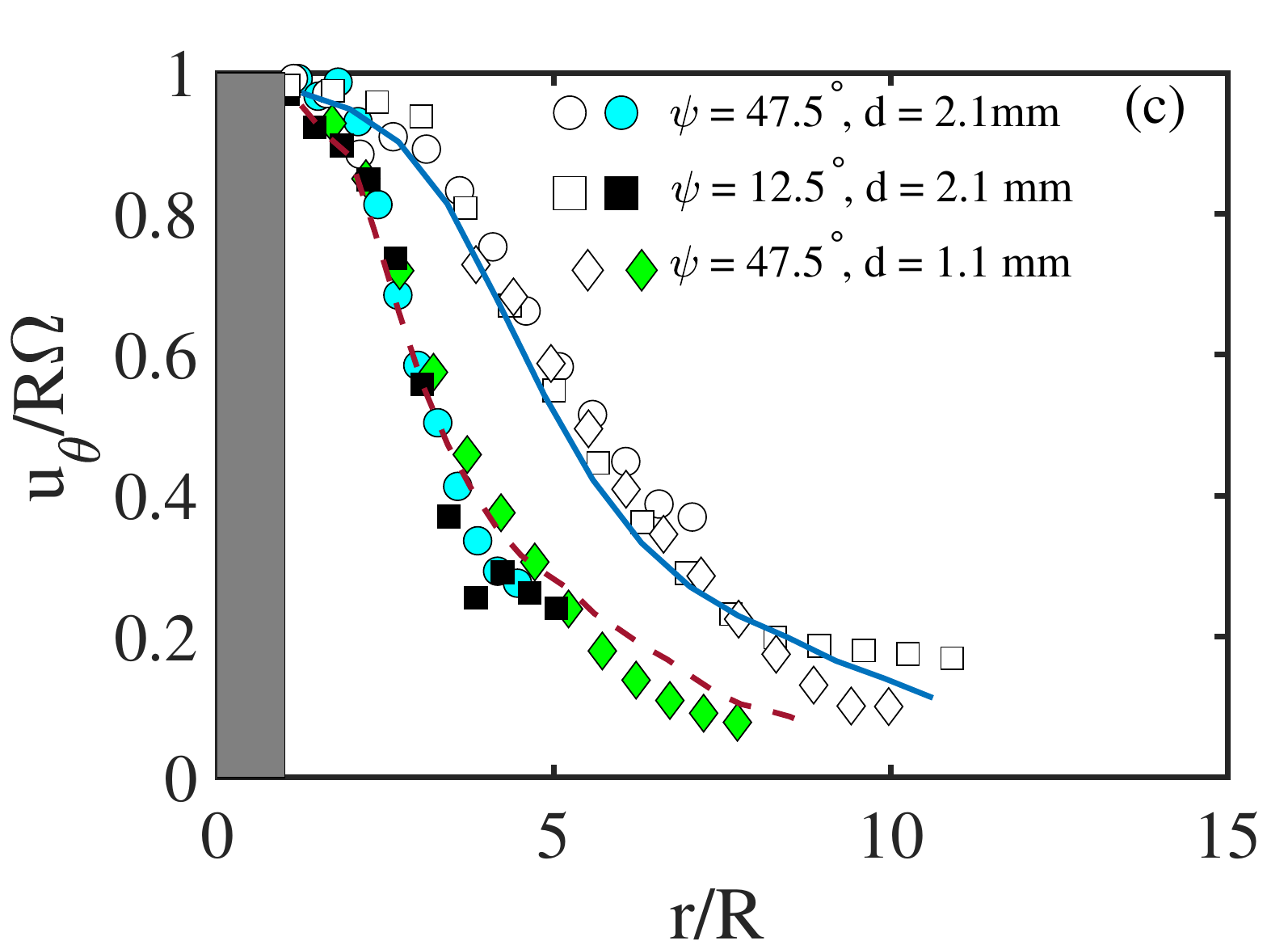}
		\caption{\small Time-averaged velocity profiles in the plane orthogonal to the direction of swimming for a Newtonian fluid at the head (a) and at the tail (b) of a swimmer with $\psi = 47.5^{\circ}$ and $d = 1.1 mm$. (c) The averaged normalized rotational velocity ($u_{\theta}/{R\Omega}$) of the fluid around different swimmers \KSC{as a function of radial distance $r/R$} in Newtonian and the viscoelastic fluids at $\Omega = 0.7~$Hz. \KSC{Here $r$ is defined as $r = \sqrt{Y^{2}+Z^{2}}$}. Empty and filled symbols show the velocity profile at the tail and the head, respectively. The dashed and continuous lines correspond to the flow around the Newtonian fluid at the head and the tail of the swimmer, respectively. }
		\label{fig4}
\end{figure}
To better understand the nature of swimming speed hindrance or enhancement as a result of swimmer's tail thickness and/or pitch angle, we probe the detailed form of the flow field around the swimmer in the viscoelastic fluid and compare it with the Newtonian fluid. \par 
First, we present the time-averaged flow fields measured for viscoelastic and Newtonian fluids in the plane orthogonal to the direction of motion. Fig.~\ref{fig4} shows two representative time-averaged normalized rotational velocity fields ($u_{\theta}/{R\Omega}$) around the swimmer in a Newtonian fluid at the swimmer head (Fig.~\ref{fig4} (a)) and at a location 1/3 of the tail length away from the tip of the tail (Fig.~\ref{fig4} (b)). Fig.~\ref{fig4} (c) shows the averaged rotational velocity of the fluid around the swimmer for three different swimmers in the viscoelastic as well \KSC {as} the Newtonian fluid (curves). The rotational velocity profiles of Fig.~\ref{fig4}(c) are obtained by averaging over the circumference of the swimmer except for the area that is dark behind the swimmer. Note the dark region is a shadow behind the swimmer, which is generated by the laser passing through the swimmer. Two observations are worth noting here. First, remarkably, the velocity fields around Newtonian and viscoelastic fluids are similar at both the head and the tail for various pitch angles and tail thicknesses. Therefore, the rotational flow around the swimmer is not affected by the rheology of the fluid and/or the swimmer properties. Secondly, the rotational flow extends farther away from the swimmer surface at the tail compared to the head.\par
\begin{figure}
	\centering
\includegraphics[width=1\textwidth]{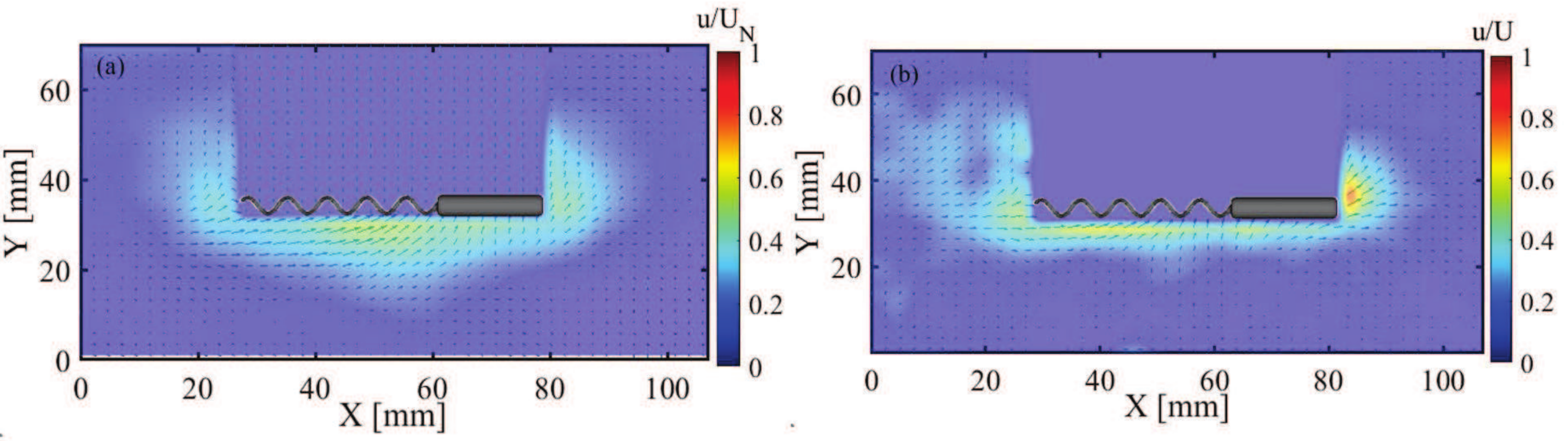}
	\caption{\small 2D velocity map of the fluid around a helical swimmer with $\psi  = 49.5^{\circ}$ and $d = 1.1$mm rotating at $\Omega = 0.1$ Hz in (a) Newtonian and (b) viscoelastic fluid. The color bar indicates the fluid velocity normalized by the swimming speed of the swimmer. }
	\label{fig5}
\end{figure}
On the other hand, Fig.~\ref{fig5} shows the time-averaged flow field around a swimmer rotating at $\Omega = $ 0.1 Hz in the longitudinal plane (X-Y plane) with a pitch angle $\psi = 49.5^{\circ}$ and the tail thickness $d = 1.1$mm in the Newtonian and viscoelastic fluids. \KSC{In Fig.~\ref{fig5}, u refers to the velocity magnitude of the fluid around the swimmer}. The time-averaged velocity profiles are obtained over several rotation cycles. We recall that the normalized swimming speed for this swimmer is around unity. In general, the flow fields in the Newtonian and the viscoelastic fluids are very similar.\par

\begin{figure}
	\centering
	\includegraphics[width=1\textwidth]{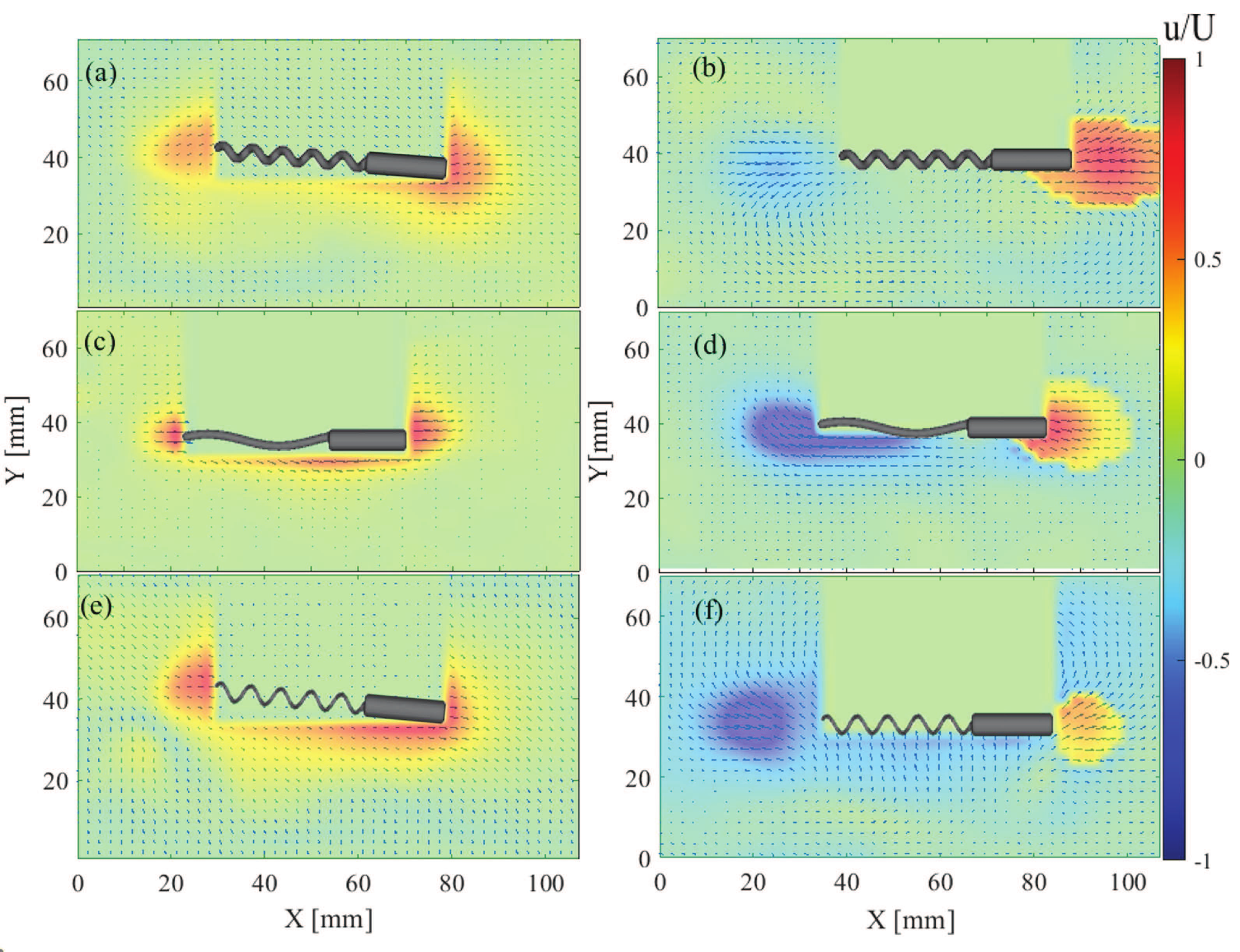}	
\caption{\small Time-averaged velocity field around a swimmer in the Newtonian fluid (a,c,e) and the viscoelastic fluid (b,d,f). The swimmer pitch angle and tail thickness ($\psi$,$d$) are ($47.5^{\circ}$, $2.1~mm$) for (a,b), ($12.5^{\circ}$, $2.1~mm$) for (c,d), and ($47.5^{\circ}$, $1.1~mm$) in (e,f).
}
	\label{fig6}
\end{figure}

Building \KSC {upon} these results, the flow field around the swimmer at higher $De$ numbers was resolved. Fig.~\ref{fig6} shows a series of time-averaged velocity profiles for Newtonian (to the left) and viscoelastic (to the right) fluids at a fixed rotational frequency of $\Omega = $ 0.7 Hz (or equivalently $De = 1.05$ in the viscoelastic fluid) and various pitch angles and tail thicknesses. For the swimmer with $\psi = 49.5 ^{\circ}$ and $d = 2.1 mm$ in the Newtonian fluid (Fig.~\ref{fig6}(a)), the front-back flow is similar to that of Fig.~\ref{fig5}(a). However, interestingly, for the same conditions in the viscoelastic fluid, we observe a strong front-back flow asymmetry that is characterized by \KSC{the }formation of a negative wake \KSC{downstream of the swimmer's body} (Fig.~\ref{fig6}(b)). To the best of our knowledge, this is the first report of its kind on the formation of a strong negative wake downstream of a helical swimmer in viscoelastic fluids.\par
At a smaller pitch angle ($\psi = 12.5^{\circ}$), for which $U/U_{N}<1$, the same front-back flow asymmetry is observed in the viscoelastic fluid (see Fig.~\ref{fig6}(d)). Evidently, as the pitch angle decreases, the backward flow \KSC{downstream of the swimmer's body} becomes stronger in the viscoelastic fluid (compare Fig.~\ref{fig6}(b) with Fig.~\ref{fig6}(d)). Furthermore, at a smaller tail thickness $d = 1.1 mm$ (Fig.~\ref{fig6}(e,f)), the Newtonian flow around the swimmer does not change appreciably compared to other \KSC{Newtonian  cases shown in Fig.~\ref{fig6}(a,c)}. However, a direct comparison between flow fields in the viscoelastic fluid (cf., Fig.~\ref{fig6}(b) and Fig.~\ref{fig6}(f)) reveals that as the tail thickness decreases, the negative wake \KSC{downstream of the swimmer's body} becomes stronger. Sample representative movies used for the above PIV analysis are included as supplementary materials.\par  

In summary, our experiments indicate that the viscoelastic flow around the helical swimmer is accompanied by a striking front-back flow asymmetry at high Deborah numbers. This flow asymmetry is characterized by formation of a negative wake in the rear of the swimmer. The strength of the negative wake is inversely proportional to the normalized swimming speed. How can we understand the swimming dynamics and the intertwined effects of swimmer shape and viscoelasticity in the context of this flow asymmetry? To shed more light on the above experimental observations, we performed 3D numerical simulations of helical swimming in a FENE-P fluid model for conditions that are matched to those of experiments.  \par

\subsection{Numerical Results}
Based on the experimental results of Fig.~\ref{fig6}, the viscoelastic flow around the swimmer is very complex with a strong extensional flow in the wake of the swimmer. Therefore, for simulations we use a FENE-P fluid model, which is known to accurately describe the extensional rheology of viscoelastic fluids~\citep{Ent97}. The simulations are performed using the exact swimmer dimensions, a viscosity ratio $\beta = 0.9$ and a finite extensibility $L^{2}\approx 9$. \par

\begin{figure}
		\centering
		\includegraphics[width=1\textwidth]{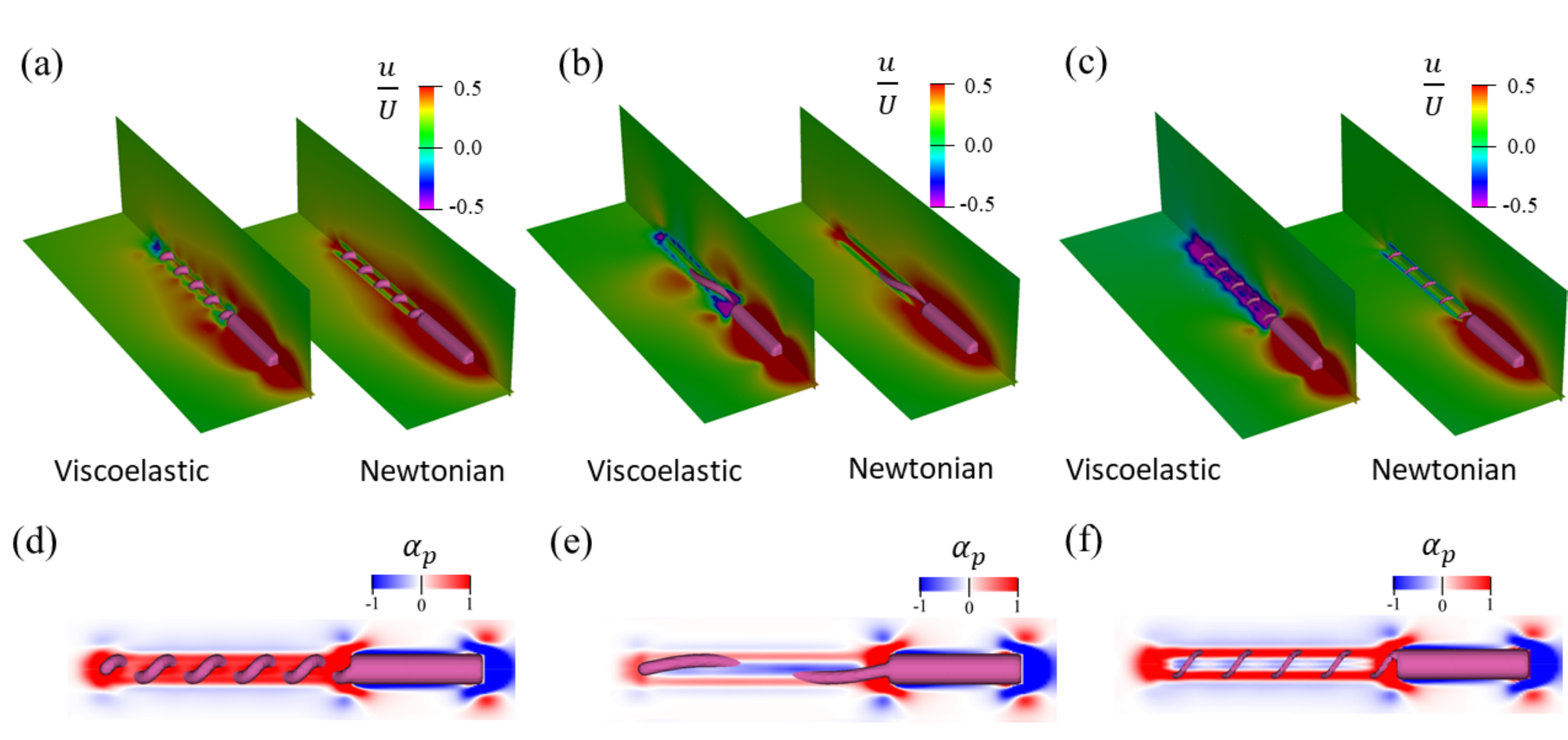}
		\caption{(a-c) The relative mean streamwise velocity of a swimmer in Newtonian and FENE-P viscoelastic fluids with $\beta=\mu_s/(\mu_s+\mu_p)=0.9$, $De=1.05$ for   ($47.5^{\circ}$,$2.1 mm$) in (a) and ($12.5^{\circ}$,$2.1mm$) in (b) and ($47.5^{\circ}$,$1.1 mm$) in (c). (d-f) the mean contribution of the viscoelastic stress tensor in the streamwise direction ($\overline{\frac{\partial{\tau_e}_{i1}}{\partial x_i}}=L_h \frac{1-\beta }{\beta De}\alpha_p$) of the corresponding a-c cases, respectively. }
		\label{fig7}
	\end{figure}

Fig.~\ref{fig7}(a-c) show averaged steady state flow velocity for three representative cases with different helix pitch angles and thicknesses in the Newtonian and the viscoelastic fluids that correspond to experimental conditions of Fig.~\ref{fig6}. The most substantial changes are observed at the helix region. First, consistent with our experiments, a negative wake \KSC{downstream of the swimmer's body} is observed in simulations of the FENE-P model. Secondly, increasing the tail thickness at a fixed tail pitch angle, gives rise to a weaker negative wake in simulations (cf., Fig.~\ref{fig7}(a) and Fig.~\ref{fig7}(b)). Additionally, decreasing the pitch angle at a fixed tail thickness gives rise to a stronger negative wake behind the swimmer (cf., Fig.~\ref{fig7}(b) and Fig.~\ref{fig7}(c)). These trends are consistent with our experimental observations. \par
\vspace{-0.5cm}

\section{Discussion}
To better understand the intertwined impacts of swimmer shape, viscoelasticity and front-back flow asymmetry in viscoelastic fluids, we discuss our results using the following approximate force balance. If the flow is approximated as linear Stokes problem, and the viscoelastic stress $\bm{\tau}_e$ in Eq.~{2.1}(a) is known as \KSC {an} a \emph{priori} volumetric force from the direct numerical simulation, we can decompose the total axial (streamwise) force on the swimmer into two terms: The first term is the net axial Newtonian force caused by translational motion (with the velocity of $U$) and rotation of the helix (with the rotational velocity of $\Omega$), which can be approximated as~\citep{GRAY55}: 
\begin{equation}
F_{s}= (A_{0}+A_{H}) U -B_{H} \Omega.
\end{equation}
Here $A_0$ is a positive resistance coefficient and is a function of the size and shape of the swimmer body and the fluid viscosity. The coefficients $A_H$ and $B_H$ are related to the helix shape and can be approximated using resistive-force theory~\citep{GRAY55} as: 
\begin{equation}
A_{H}\approx (c_{\|} \cos^{2} \psi +c_{\perp} \sin^{2} \psi) L_{H},
\end{equation}
\begin{equation}
B_{H}\approx(c_{\perp}-c_{\\})\sin \psi \cos \psi R L_{H},
\end{equation}
with $L_H$ being the length of helix and $c_{\perp}=2c_{\|}=4\mu_s \pi/\ln(2/\epsilon)$, where $\epsilon \ll 1$ is the aspect ratio of the helix. \par

The second term in the net axial force is caused by the viscoelastic stress tensor in the streamwise direction, and can be written as the volume integral: $F_e=-\int_V \frac{\partial{\tau_e}_{i1}}{\partial x_i} dV$. At low Reynolds numbers, $F_{s}$ and $F_{e}$ must balance and the force free-swimming transitional velocity can be expressed as:
\begin{equation}
U=\frac{B_{H}}{A_{0}+A_{H}}\Omega-\frac{1}{A_{0}+A_{H}} F_{e}. 
\end{equation}
The first term is approximately similar to the Newtonian swimming velocity $U_N=\frac{B_H}{A_0+A_H}\Omega$ \citep{lauga2020fluid} and thus, the normalized swimming speed can be written as:
\begin{equation}
\frac{U}{U_{N}}\,=\,1\,+\,\frac{1}{{B_{H} \Omega}}\int_V \frac{\partial{\tau_{e}}_{i1}}{\partial x_{i}} dV.
\end{equation}
\par  
Fig.~\ref{fig7}(d-f) show the contour plot of the above volumetric integral for the viscoelastic cases corresponding to  Fig.~\ref{fig7}(a-c). While the viscoelastic effects are almost similar near the head and around the swimmer for different cases, there is a substantial difference in the interior of the helix. For the swimmer with $d=2.1 mm$ and $\psi=47.5^{\circ}$, the whole region inside the helix is encapsulated by viscoelastic forces that contribute to the thrust. This prediction leads to \KSC {a} swimming speed enhancement, which is consistent with the experimental observations. For other swimmers that correspond to Fig.~\ref{fig7}(a,c), the viscoelastic stresses at the inner region of the helix are predicted to contribute into drag enhancement, which is again consistent with the experimental observations for such cases. Therefore, the above simple force balance argument can explain the swimming dynamics in our experiments based on the contribution of the viscoelastic forces in the region interior to the helix. \par

Our results suggest that the non-local effects of the viscoelatsic stress tensor inside the helix region play a key role in determining the dynamics of swimming in viscoelastic fluids. The strong rotational shear flow of viscoelastic fluid around the swimmer body gives rise to a hoop stress that develops due to stretching of the polymer molecules along the curved streamlines, and causes a strong extensional flow (or normal stress differences) in the interior region of the helix along the streamwise direction. The extensional flow in the interior of the helix extends to the back of the swimmer and manifests itself as a front-back flow asymmetry accompanied by a strong negative wake \KSC{downstream of the swimmer's body}. \KSC{However, the results suggest that the front-back flow asymmetry (or the strong negative wake) by itself is not a discriminating metric to rationalize the effects of viscoelasticity on swimming speed. Instead, the progressive engagement of the core flow in the interior region of the helix is the key to better understanding of the impact of viscoelastic forces on swimming dynamics. This mechanism is principally related to local viscoelastic effects near the rotating helix and is inherently distinct from what has been shown theoretically for a snowman swimmer,  a dumbbell composed of two spheres of different diameters \citep{pak2012micropropulsion,angeles2021front}. } In a recent paper, \cite{Ang2021} showed that the swimming dynamics of helical swimmers in viscoelastic fluids depend strongly on the front-back geometrical asymmetry; i.e., the difference between \KSC{the} head and the helix diameter. The effects of front-back geometrical asymmetry were explained in the context of snowman-like effects \KSC {(i.e., effects due to asymmetry between the diameter of the head and the tail of the swimmer)}~\citep{Ang2021}. In our paper, the diameters of the head and helix are the same, and therefore snowman-like effects should be minimal. Instead, we think that the contribution of the normal stresses inside the helix is the key parameter in the enhancement or diminution of the swimming speed. These stresses are strongly controlled by the properties of the polymer solution and the finite extensibility of the polymer chains.  \par

\section{Conclusion}
In summary, we have demonstrated that increasing the tail pitch angle and/or the tail thickness of a helical swimmer gives rise to \KSC{a} swimming speed enhancement in a viscoelastic fluid. More importantly, we presented the first evidence for formation of a front-back flow asymmetry around a swimmer, which is characterized by formation of a negative wake downstream of the helical swimmer in the viscoelastic fluid. Our simulations  predict a similar front-back flow asymmetry observed in experiments. Furthermore, using a simple force balance argument, we showed that the contribution of the viscoelastic forces in the interior region of the helix are key in controlling the swimming dynamics. These streamwise viscoelastic forces are generated by the hoop stress and are at the origin of the front-back flow asymmetry (or the negative wake \KSC{downstream of the swimmer's body}) in experiments. \KSC{ Despite the negative wake being a signature of swimming in viscoelastic fluids, it is not directly correlated with the enhancement or reduction of swimming speed. Instead, the gradual contribution of the core flow region inside the helix and the net effect in immediate region of the helix wake are recognized as the main characteristic of the viscoelastic cases with enhanced swimming velocity.} \par

\KSC{Finally, in the current study the helix is actuated via an external torque, while swimming of flagellated bacteria is torque-free and the bacteria body always rotates in an opposite direction to the rotating helix \citep{lauga2020fluid}. Further research is required to address how the swimming performance is affected by these effects, namely the torque-free swimming and the counter rotations of the helix and the cell body.}

\vspace{-0.5cm}
\section{Acknowledgements} 
The authors acknowledge the Florida State University Research Computing Center for the computational resources on which these simulations were carried out and Florida State University Council for Research and Creativity for partially supporting this work.

\vspace{-0.5cm}

\bibliographystyle{jfm}
\bibliography{jfm-instructions}

\end{document}